\documentstyle[graphicx,floats,prl,aps]{revtex}
\def \BE {\begin{equation}}
\def \EE {\end{equation}}
\def \BEAH {\begin{eqnarray*}}
\def \EEAH {\end{eqnarray*}}
\def \BEA {\begin{eqnarray}}
\def \EEA {\end{eqnarray}}
\def \BDM {\begin{displaymath}}
\def \EDM {\end{displaymath}}
\begin{document}
\draft
\title{Uniformly accelerated sources in electromagnetism and gravity}
\author{V. Pravda\footnote{Email address: {\tt pravda@otokar.troja.mff.cuni.cz} }
\ and A. Pravdov\' a \footnote{Email address: {\tt pravdova@otokar.troja.mff.cuni.cz}}}

\address{Department of Theoretical Physics,
         Faculty of Mathematics and Physics,
         Charles University, \\
         V Hole\v sovi\v ck\i ach  2,
         180 00 Prague 8,
         Czech Republic}
\date{June 1998}

\maketitle

\begin{abstract}
Electromagnetic field produced by magnetic
multipoles in hyperbolic motion is derived and compared with electromagnetic field produced
by electric multipoles in hyperbolic motion. The resulting fields are
related by duality symmetry. Radiative properties of these
solutions are demonstrated.  In the second part an analogous, uniformly accelerated source
of gravitational radiation is studied, within exact Einstein's theory.  Radiative characteristics
of the corresponding solution as flux of the radiation and the total mass-energy
 of the system are calculated and graphically illustrated.
\end{abstract}

\section{Introduction}
Because of nonlinearity of
Einstein's equations it is difficult to find their realistic exact
radiative solutions. If we want to describe gravitational
radiation from a given realistic source, we have to use various
approximation methods, but then doubts arise if such
approximate solutions correspond to some exact solutions.

It is easier to find exact solutions if
we assume some symmetries. If we consider axially symmetric
space-times which are asymptotically flat at least locally, the
only second allowable symmetry that does not exclude radiation is
the  boost symmetry (see \cite{JBS}, \cite{AJB}). Boost-rotation symmetric solutions describe
"uniformly accelerated particles" of various kinds and they
contain gravitational radiation. These exact solutions help us
to understand  properties of gravitational radiation and
they can also be used  as tests of various approximation methods
or numerical computations. In fact some specific boost-rotation symmetric
solutions were used already as test beds in numerical relativity
 \cite{Alcubierre}.

Gravitational radiation of uniformly accelerated particles and
electromagnetic radiation of the analogous system of charges have
some similar properties. In the first part of this paper we present
new solutions of Maxwell equations describing uniformly
accelerated magnetic multipoles, we compare them with solutions
describing  uniformly accelerated electric multipoles found in
\cite{Muschall}
(the case of monopole having been studied first by Born
(1909))
and we analyze their radiative properties.

In the second part we turn  the to boost-rotation
symmetric solutions in  general relativity. We analyze outgoing
gravitational radiation from uniformly accelerated
particles described by specific solutions and we calculate the mass
decrease of the radiative system caused by  energy carried
out from the system by gravitational radiation.

\section{Electromagnetic fields and radiation patterns from magnetic
multipoles in hyperbolic motion}

\subsection{Construction of the solution}
Assume that a particle with dipole magnetic moment $m_0$ moves
with a uniform acceleration $\alpha^{-1}>0$ along the $z$-axis of
cylindrical coordinates $(t,\rho,\phi,z)$ in Minkowski space-time.
Its worldline is the hyperbola
\BE
\rho=0, \quad  z=\sqrt{\alpha^2+t^2}. \label{svetocar}
\EE
If we want to calculate field produced by this particle we have to
know corresponding four-current $J^{\alpha}$.

Electromagnetic field produced by magnetic dipole at rest is
\BDM
\vec A={\frac {{\vec m_0} \times {\vec r}}{{r}^{3}}}.
\EDM
Now we can use relation $\Delta \vec A=-4 \pi \vec  \jmath \,$ and calculate thus the
corresponding current.
If we apply $\partial_z$  on equation
\mbox{$\Delta({\frac{1}{r}})=-4 \pi \delta(x)\delta(y)\delta(z)$}, we
obtain
$\Delta( {z}/{r}^{3})=4 \pi \delta(x)\delta(y)\delta'(z)$.
We put $z$-axis in \mbox{$m_0$-direction $\Rightarrow \vec m_0=[0,0,m_0]$},
$\vec m_0 \times \vec r = [-m_0 y,m_0 x,0]$
and we have
\BE
\vec \jmath_{\rm{rest}}={\it m_0}\,\delta(z)[\delta(x) \delta'(y),-\delta'(x)
\delta(y),0].
\EE
A  particle which is moving with  velocity $v$ with respect to a coordinate
system $S$
and has a magnetic dipole $m_0$ in its rest frame, in the frame $S$ has a magnetic
 dipole $m=m_0 \sqrt{1-v^2}$.
In the case of  uniformly accelerated magnetic dipole we have
$
\sqrt{1-{v^2}}={\alpha}/{\sqrt {\alpha^2+t^2}}
$
so the four-current of the uniformly accelerate magnetic dipole
is
\BE
J^\alpha={\frac {{\it m_0}\,\alpha\,\delta(z-\sqrt {{\alpha}^{2}+{t}^{2
}})}{\sqrt
{{\alpha}^{2}+{t}^{2}}}} \ [0,\delta(x) \delta'(y),-\delta'(x)
\delta(y),0]
\ .
\label{acc4cur}
\EE
Corresponding four-potential can be found by using the standard relation
\BE
A^\alpha =  \int \! {\frac{\delta(t'+ {{ |\vec x- \vec x'|
}-t})}
{  |\vec x- \vec x'|  }} J^\alpha \, d^3 x' dt'  .
\label{AintJ}
\EE
After quite a long calculation (details in \cite{Vojdok}) we obtain the result
 which is most suitably
 expressed in cylindrical coordinates.
We have  $A^t=A^z=A^\rho=0$ and
\BE
A^\phi =  \frac{ 4 m_0 \alpha(\alpha^2+\rho^2+z^2-t^2)}{\xi^3} \, ,
\quad {\rm where} \quad \quad
 \xi = \sqrt{4 \rho^2 \alpha^2 + (\alpha^2+ t^2 - z^2 -\rho^2)^2}
\, ,
\EE
but we will work in ortho{\it normal} tetrad basis defined by
cylindrical coordinates;  in this basis $A^\alpha$ has
non-zero component $A^{(\phi)}=\rho A^\phi$.
With the help of
$
\vec B = \nabla \times \vec A \ , \quad \vec E = -\partial_t \vec A.
$
we obtain $E^{(\rho)}=E^{(z)}=B^{(\phi)}=0$ and
\BEA
B^{(\rho)} &=& \enspace  \ 8 m_0 \rho z \partial_\alpha \left ({\frac {{\alpha}^{2}
}{{\xi}^{3}}}\right ) \ ,  \nonumber  \\
B^{(z)} &=& - 4 m_0  \partial_\alpha \left ({\frac {{\alpha}^{2}
\left ({\alpha}^{2}+{t}^{2}+{\rho}^{2}-{z}^{2}\right )}{{\xi}^{3}
}}\right )  \ , \quad \\
E^{(\phi)} &=& -  8 m_0 \rho t \partial_\alpha \left ({\frac {{\alpha}^{2}
}{{\xi}^{3}}}\right ) . \nonumber
\EEA
The field corresponding to a uniformly accelerated $2^l$-pole is
given in terms of $(l-1)^th$ derivatives with respect to the
parameter $\alpha$ of the dipole fields. For detailed consideration in the
case of electric multipoles see \cite{Muschall}  (it is based on the way how dipole is constructed from
monopoles etc., and on the fact that Maxwell equations are
linear). In the case of magnetic multipoles we find
\begin{center}
\BEA
B^{(\rho)}_{(l)}&=& \enspace  \ 8 \frac{{\cal M}_l (2l-1)!!}{l!^2} \rho z \partial_\alpha^{(l)} \left ({\frac {{\alpha}^{2}
}{{\xi}^{3}}}\right ) \ ,  \nonumber \\
B^{(z)}_{(l)} &=& - 4 \frac{{\cal M}_l (2l-1)!!}{l!^2}  \partial_\alpha^{(l)} \left ({\frac {{\alpha}^{2}
\left ({\alpha}^{2}+{t}^{2}+{\rho}^{2}-{z}^{2}\right )}{{\xi}^{3}
}}\right )  \ ,  \label{Fmgmulti} \\
E^{(\phi)}_{(l)}&=& -  8 \frac{{\cal M}_l (2l-1)!!}{l!^2} \rho t \partial_\alpha^{(l)} \left ({\frac {{\alpha}^{2}
}{{\xi}^{3}}}\right ). \nonumber
\EEA
\end{center}
According to \cite{Muschall} we define ${\cal M}_l$ as the only one independent component of the
corresponding multipole tensor up to a constant $l!^2/(2l-1)!!$.
The field corresponding to uniformly accelerated electric multipoles ${\cal E}_l$
reads (see \cite{Muschall})
\BEA
E^{(\rho)}_{(l)} &=&  \enspace  \ 8 \frac{{\cal E}_l (2l-1)!!}{l!^2} \rho z \partial_\alpha^{(l)} \left ({\frac {{\alpha}^{2}
}{{\xi}^{3}}}\right ) \ ,  \quad \nonumber \\
E^{(z)}_{(l)} &=& -4 \frac{{\cal E}_l (2l-1)!!}{l!^2}  \partial_\alpha^{(l)} \left ({\frac {{\alpha}^{2}
\left ({\alpha}^{2}+{t}^{2}+{\rho}^{2}-{z}^{2}\right )}{{\xi}^{3}
}}\right )  \ ,   \label{Felmulti} \\
B^{(\phi)}_{(l)} &=&    \enspace \ 8 \frac{{\cal E}_l (2l-1)!!}{l!^2} \rho t \partial_\alpha^{(l)} \left ({\frac {{\alpha}^{2}
}{{\xi}^{3}}}\right ). \nonumber
\EEA
We see that the field given by (\ref{Fmgmulti}) can be obtained from that given
by (\ref{Felmulti})
by simple transformation
\BE
\vec B \rightarrow -\vec E, \quad \vec E \rightarrow \vec B \ ,
\quad {\cal E}_l \rightarrow {\cal M}_l .
\label{dual1}
\EE
It is a special case of duality symmetry (see for example \cite{duality})
which is well known in the
vacuum case. To keep this symmetry in presence of charges, magnetic
sources have to be introduced.
Now the duality symmetry of Maxwell equations
$(\vec E + i \vec B) \rightarrow e^{i\Phi}(\vec E + i \vec B)$
is restored if we also rotate the electric and magnetic charges
$(q+ig) \rightarrow e^{i\Phi}(q+ig)$. We see that  (\ref{dual1})
is a special case of duality symmetry in the presence of charges.

The simplest example of fields (\ref{Fmgmulti}) and (\ref{Felmulti})
is the field of uniformly accelerated electric monopole (Born's solution).
This field  has been
studied in some basic works (Pauli 1918; Laue 1919) which considered the field
as non-radiative. Even now some authors consider this solution as
non-radiative \cite{Singal}.
 This is in contradiction with expression
according to which an accelerated charge radiates energy with the
rate $(2/3) e^2 {\dot v}^2$. If we expand the field strengths of
Born's solution in powers of $r^{-1} \ (r^2=\rho^2+z^2)$, with the
fixed time $t$, we find $\vec E \sim r^{-4}$, $\vec B \sim r^{-5}$
and thus Poynting vector $\vec S \sim r^{-9}$. In Figure 1 we
can see that the quantities determining the field have the
character of a pulse and it is therefore understandable  why we find
non-radiative Poynting vector $\vec S \sim r^{-9}$ when going
to spatial infinity and therefore passing through a pulse. In the
next part we will see that $\vec S \sim r^{-2}$ when travelling
with the pulse with the velocity of light; then all fields
(\ref{Fmgmulti}) and (\ref{Felmulti}) have radiative character.
\begin{figure}
\begin{center}
\includegraphics*[height=3in]{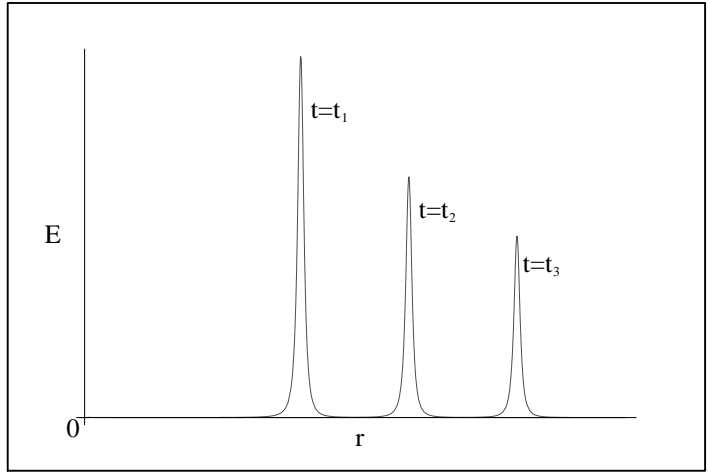}
\end{center}
\caption{Pulse generated by uniformly accelerated electric charge ($E=|\vec E| $).}
\label{fig:pulse}
\end{figure}
\subsection{Asymptotic behaviour and radiation properties}
We will express the field components (\ref{Fmgmulti}) in terms of spherical coordinates
$(r,\theta,\varphi)$ and of the retarded time of the origin $u=t-r$
and expand them in $r^{-1}$ with $u$, $\theta$, $\phi$ fixed. Neglecting
the terms $O(r^{-2})$ we find
\BEA
B^{(\rho)} &=& \enspace  \ {\cal M}_l G_l \sin \theta \cos \theta \frac{1}{r} \ ,
\nonumber \\
B^{(z)} &=& -{\cal M}_l G_l \sin^2 \theta \frac{1}{r}   \ ,   \label{leading} \\
E^{(\phi)} &=& -{\cal M}_l G_l \sin \theta  \frac{1}{r} \ , \nonumber
\EEA
where
\BE
G_l=\frac{(2l-1)!!}{l!^2} \frac{\partial^{(l)}}{\partial \alpha}
\frac{\alpha^2}{(u^2+\alpha^2 \sin^2 \theta )^{(3/2)}}\ .\label{Gl}
\EE
From (\ref{leading}) we can calculate the
leading term of the radial component of the Poynting vector
$\vec S=(1/4 \pi) \vec E \times \vec B$.
Considering a particle with an arbitrary structure of electric and
magnetic multipoles, we obtain
\BE
S_r=\frac{\sin^2 \theta}{4 \pi r^2} \sum_{l,l'}
({\cal E}_l {\cal E}_{l'}+{\cal M}_l {\cal M}_{l'}) G_l G_{l'} +
O(r^{-3}) \ .
\EE
Introducing the true retarded time $u^*$ of the particle with the
help of the relation
\BE
t-u^*=|\vec r - \vec r^{\, *}| \ ,
\EE
where $(t,\vec r)$ are the coordinates of an observation event
and $(u^*,\vec r^{\, *})$ are the coordinates of an emission event, we
can write
\BE
u=u^*-\sqrt{{u^*}^2+\alpha^2} \cos \theta + O(r^{-1})
\label{defu} \ .
\EE
Radial flux emitted at $u^*=0$ reads
\BE
S_r=\frac{\sin^2 \theta}{4 \pi r^2} \sum_{l,l'}
 \frac{(2l-1)!! (2l'-1)!!({\cal E}_l {\cal E}_{l'}+{\cal M}_l {\cal M}_{l'})}{l!^2
 l'!^2}  \frac{V_l V_{l'}}{\alpha^{l+l'+2}} \ ,
\EE
where $V_l$ can be obtained by substituting (\ref{defu}) and $u^*=0$
in
\BDM
\alpha^{l+1} \, \frac{\partial^{(l)}}{\partial \alpha}
\frac{\alpha^2}{(u^2+\alpha^2 \sin^2 \theta )^{(3/2)}} \ ;
\EDM
thus $V_0=1$, $V_1=3 \cos^2 \theta -1$, $V_2=15 \cos^4 \theta-15 \cos^2
\theta+2$, $\dots$. \\
The total radiated power  $R=2 \pi \int_{0}^{\pi} {\lim\limits_{r \to
\infty}}
(r^2 S_r) \sin \theta \, {\rm d} \theta$ can be expressed in the
form
\BE
R=\sum_{l,l'} R^{(l,l')} \frac{{\cal E}_l {\cal E}_{l'}+{\cal M}_l {\cal
M}_{l'}}{{\alpha^{l+l'+2}}} \label{power} \ ,
\EE
with
\BE
R^{(l,l')}=\frac{(2l-1)!! (2l'-1)!!}{2l!^2 l'!^2} \int_{0}^\pi V_l
V_{l'} \sin^3 \theta \, {\rm d} \theta \ ;
\EE
for all ${\cal M}_l=0$ (\ref{power}) is
 identical with (23) in \cite{Muschall}. As in \cite{Muschall} we have
 $R^{(0,0)}=2/3$, $R^{(0,1)}=-4/15$, $R^{(1,1)}=8/21$ and so on. Due to the boost
 symmetry, the particles radiate out energy with a constant rate
 independent of $u^*$, and consequently with the same rate as at
 the turning point $u^*=0$. For the detailed consideration, see
 \cite{bicakBS}.

In terms of the electromagnetic field tensor, $F_{\mu \nu }$, we obtain, in
coordinates $u$, $r$, $\theta$, $\phi$, the leading (in $r^{-1}$) radiative
terms of a general boost-rotation symmetric electromagnetic field in the form:
\BEA
F_{u\theta}&=&X=\frac{\epsilon (w)}{u^2} \ , \nonumber \\
F_{u\phi}&=&Y\sin\theta=\frac{\beta(w)\sin\theta}{u^2} \nonumber  \\
&& {\rm with} \ \  w=\frac{\sin\theta}{u} \ , \nonumber
\EEA
in which $X$ and $Y$ correspond to the so called news functions of the
system, using the
 general-relativistic terminology (see \cite{AJB}).
(In cartesian coordinates these
terms imply terms $\sim r^{-1}$.)
For a uniformly accelerated
electric monopole (i.e. Born's solution) we obtain
\BEA
{\epsilon}(w)&=&\frac{{\cal E}_0 \alpha^2 w}{(1+\alpha^2 w^2)^{3/2}}, \nonumber \\
{\beta}(w)&=&0. \nonumber
\EEA
For a uniformly accelerated
electric dipole  we get
\BEA
{ \epsilon}(w)&=&\frac{{\cal E}_1 \alpha w(2-\alpha^2 w^2)}{(1+\alpha^2 w^2)^{5/2}} \ , \nonumber \\
{ \beta}(w)&=&0 \ , \nonumber
\EEA
 whereas for
magnetic dipole  one finds
\BEA
{ \epsilon}(w)&=&0 \ , \nonumber \\
 {\beta}(w)&=&- \frac{{\cal M}_1 \alpha w(2-\alpha^2 w^2)}{(1+\alpha^2 w^2)^{5/2}} \ . \nonumber
\EEA
For a general electric $2^l$-pole
\BEA
\epsilon(w)&=&{\cal E}_l (G_l u^3)w \ , \nonumber \\
{ \beta}(w)&=&0 \ , \nonumber
\EEA
and for a magnetic $2^l$-pole
\BEA
{ \epsilon}(w)&=&0 \ , \nonumber \\
\beta(w)&=&-{\cal M}_l (G_l u^3)w  \ , \nonumber
\EEA
with $G_l$ given by (\ref{Gl}).
\section{
An example of radiative particles in general relativity}
The Bonnor-Swaminarayan solution is a boost-rotation symmetric
solution of Einstein's equations. It describes the gravitational field
around a finite number of monopole Curzon-Chazy particles
uniformly accelerated in opposite directions. The acceleration force
is caused  by gravitational interaction among particles or by
nodal singularities. The most interesting BS-solution
contains two pairs of particles, there is one particle with positive
and one with negative mass in each pair.

BS metric in cylindrical coordinates $t$, $\rho$, $z$ and $\phi$ reads
\BE
ds^2=-e^\lambda d\rho^2-\rho^2 e^{-\mu}d\phi^2
     +\frac{1}{z^2-t^2}\Bigl\{ (z^2 e^\mu-t^2 e^\lambda)dt^2
                            -(z^2 e^\lambda-t^2 e^\mu)dz^2
                         +2zt(e^\lambda -e^\mu)dzdt\Bigr\}\ ,
                         \label{Bsmetric}
\EE
in which functions entering the metric have forms
\BEA
\mu&=&-\frac{2a_1}{R_1}-\frac{2a_2}{R_2}
     +\frac{2a_1}{h_1}+\frac{2a_2}{h_2}+\ln k\ ,\nonumber\\
\lambda&=&\frac{a_1 a_2}{(h_1-h_2)^2}f
     -\rho^2(z^2-t^2)(\frac{a_1^2}{R_1^4}+\frac{a_2^2}{R_2^4})
     +\frac{2a_1 R}{h_1 R_1}+\frac{2a_2 R}{h_2 R_2}+\ln k\ ,\nonumber\\
R &=&\frac{1}{2} (\rho^2+z^2-t^2)\ ,\\
R_i&=&\sqrt{(R-h_i)^2+2\rho^2 h_i}\ ,i=1,2\ ,\nonumber\\
f  &=&\frac{4}{R_1 R_2}\Bigl\{
\rho^2(z^2-t^2)+(R-\rho^2-h_1)(R-\rho^2-h_2)-R_1R_2\Bigr\}\ ,\nonumber
\EEA
where $a_1$, $a_2$, $h_1>0$, $h_2>0$, $k>0$ are constants.
In the case we are interested in, that is two pairs of freely
moving positive and negative particles, we have
\BE
a_1=\frac{(h_1-h_2)^2}{2h_2}\ ,\
a_2=-\frac{(h_1-h_2)^2}{2h_1}\ ,\
k  =1\ .
\EE

To examine the radiative properties and to find the news function
of this solution it is necessary to transform the metric at
first to spherical flat-space coordinates
$\{R,\ \vartheta,\ \phi\}$ by
$\rho=R \sin\vartheta$, $z=R\cos\vartheta$, $\phi=\phi$,
and flat-space retarded time $U=t-R$ (see \cite{bicakBS})
and then find a transformation to Bondi's coordinates $u$, $r$, $\theta$ and $\phi$
(for details see  \cite{bicakBS} and \cite{bondi}),
in which the metric has Bondi's form (see \cite{AJB,bondi} or
(2), (4) in \cite{bicakBS}). Bondi with collaborators discovered that
these coordinates are most suitable for studying radiating  systems.
Coordinates $u$, $\theta$ and $\phi$ are such that they are stant
and $r$ varies along  outgoing
null geodesics, i.e. light rays; the area of the surface
element $u=$const, $r=$const is $r^2\sin\theta d\theta d\phi$ .

In our case of two freely falling particles in each pair in
the limiting case of small
masses ($h_2=h$, $h_1=h+\epsilon$, $\epsilon>0$ small, $k=1$ and masses are then
$m^{(1)}=\epsilon^2 / 2h\sqrt{2h+2\epsilon}$, $m^{(2)}=-\epsilon^2 /(2h+2\epsilon)\sqrt{2h}$)
the relation between the flat coordinates $\{U,\ R,\ \vartheta,\ \phi\}$
and Bondi's coordinates $\{u,\ r,\ \theta,\ \phi\}$ is (see
\cite{bicakBS})
\BE
U=\Bigl( u+O(\epsilon^3 )\Bigr) +O(1/r)\ ,\ R=r+O(1)\ ,\ \vartheta=\theta+O(1/r)\ ,\ \phi=\phi\ .
\EE
Then the news function in Bondi's coordinates reads
\BEA
c,_u&= & \epsilon^3\frac{3u\sin^2\theta}{2h(u^2+2h\sin^2\theta)^{\frac{5}{2}}}\ ,\nonumber\\
    &= & \epsilon^3\frac{3w^2  }{2h(1  +2hw^2 )^{\frac{5}{2}}}\frac{1}{u^2}\
         {\rm{for}}\  u>0\ ,\ {\rm{where}} \ w=\frac{\sin\theta}{u}\ ,\\
    &= &-\epsilon^3\frac{3w^2  }{2h(1  +2hw^2 )^{\frac{5}{2}}}\frac{1}{u^2}\
         {\rm{for}}\  u<0\ .\nonumber
\EEA
As was shown in \cite{AJB}, a news function corresponding to
any asymptotically flat
boost-rotation symmetric solution of Einstein's equations has to
have a form $c,_u={\cal K}(w) /u^2$. Thus, for this case
\BEA
{\cal K} (w)&=& \epsilon^3\frac{3w^2  }{2h(1  +2hw^2 )^{\frac{5}{2}}}\
                  {\rm{for}}\ u>0\ ,\\
{\cal K} (w)&=&-\epsilon^3\frac{3w^2  }{2h(1  +2hw^2 )^{\frac{5}{2}}}\
                  {\rm{for}}\ u<0\ .\nonumber
\EEA
We also need  the news function to calculate the total mass of the
system and to show how this mass decreases due to the emission of
gravitational waves: 
\BEA
m&=&\frac{1}{4}\int_0^\pi (w^2{\cal K},_w),_w d\theta
     +\frac{1}{2}\int_0^\pi \frac{\lambda (w)}{w^2 u^2}d\theta\ ,\nonumber \\
     {\rm{where}} \quad \quad \quad  \lambda (w),_w&=&w^2{\cal K},_w^2
                  -\frac{1}{2w}(w^3{\cal K},_{ww}),_w\ .
\EEA
Fairly long calculations lead to the total Bondi mass of the form
\BEA
m&= &\frac{1}{4096}\frac{o^2u^3(5u^4+32u^2h+64h^2)\sqrt{2}
            (-\ln(u^2)+\ln \left |-2\sqrt{2h}\sqrt{u^2+2h}+u^2+4h\right |)}
              {h^{7/2}(u^2+2h)^{7/2}}\\
 &\ &+\frac{1}{3072}\frac{o^2(u^2+4h)(15u^4+16u^2h+32h^2)}{h^3u(u^2+2h)^3}\
          ,\ \quad \quad \quad {\rm{where}} \ o=\frac{3\epsilon^3}{2h}\ .\nonumber
\EEA
In Figure 2  we see that $m$ as a function of $u$ is
everywhere decreasing. The system does radiate gravitational
waves. \\[2mm]
\begin{figure}
\begin{center}
\includegraphics*[height=3in]{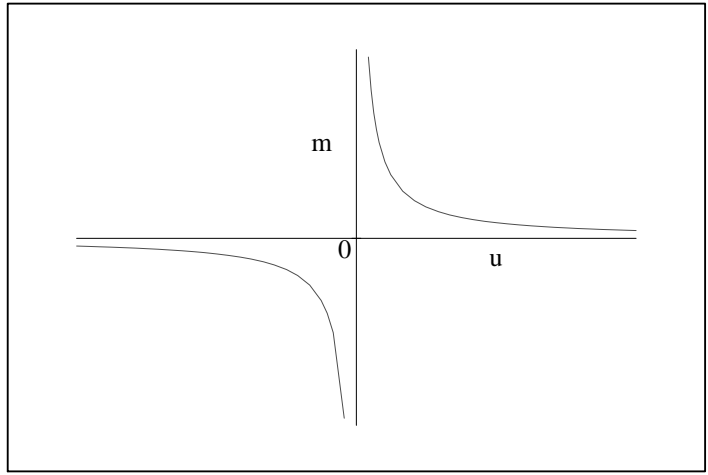}
\end{center}
\caption{Decrease of the total mass-energy of the radiating system
described by the metric (\ref{Bsmetric})}
\end{figure}
\begin{figure}
\begin{center}
\includegraphics*[height=3in]{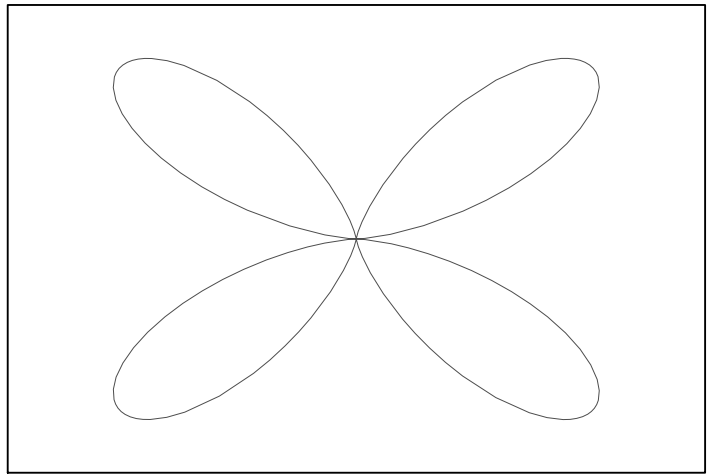}
\end{center}
\caption{Radiation pattern emitted at the turning point
by particles described by the metric (\ref{Bsmetric})}
\end{figure}
{{\bf{Acknowledgments}}}\\[2mm]
We thank J. Bi\v c\' ak for suggesting the problem and for constant
encouragement. Thanks are also due to P.~Ba\v ckovsk\' y
for stimulating discussions. \newpage

\end{document}